Statistical testing in a Linear Probability Space 18 March 2019

Christopher M Rembold MD
Cardiovascular Division
Department of Internal Medicine
University of Virginia Health System
Charlottesville, Virginia 22908-0146 USA

Running Head: Statistical testing in Linear Probability Space

Contact information:
Christopher M. Rembold, M.D.
Box 800146
Cardiovascular Division
University of Virginia Health System
Charlottesville, Virginia 22908-0146 USA
Telephone (434) 924-2825
Email: crembold@virginia.edu

Word count: text 6141, abstract 147, tables 0, figures 8

Conflicts of interests: none

All authors had access to the data and had a role in writing the manuscript (obviously since there is one author)

## ABSTRACT

Imagine that you could calculate of posttest probabilities, i.e. Bayes' theorem with simple addition. This is possible if we stop thinking of probabilities as ranging from 0 to 1.0. There is a naturally occurring linear probability space when data are transformed into the logarithm of the odds ratio ($\log_{10}$ odds). In this space, probabilities are replaced by W (Weight) where $W=\log_{10}(\text{probability}/(1–\text{probability}))$. I would like to argue the multiple benefits of performing statistical testing in a linear probability space: 1) Statistical testing is accurate in linear probability space but not in other spaces. 2) Effect size is called Impact (I) and is the difference in means between two treatments ($I=W_{mean2}–W_{mean1}$). 3) Bayes theorem is simply $W_{posttest}=W_{pretest}+I_{test}$. 4) Significance (p value) is replaced by Certainty (C) which is the W of the p value. Methods to transform data into and out of linear probability space are described.

## BACKGROUND

"Misunderstanding and misuse of statistical significance impede science" was the subtitle of Steven Goodman's 2016 perspective in *Science* (*1*). Many scientists, journals, and regulators think that a p= 0.049 suggests truth and a p=0.051 suggest falsity; this is clearly incorrect reasoning. When a single experiment with a large n has a p=0.049, a subsequent identical study has only a 50.4% chance of finding a p<0.05 (*2, 3*). This reliance on p values forms part of the "frequentist" camp of statistics. Frequentists ignore pretest likelihood at their peril (see https://xkcd.com/1132/). This use of p values, among other causes, has led to a reproducibility crisis in science (*4*). Finally, much of the non-statistically trained general population thinks the term "significance" suggests importance rather than certainty.

Prof. Goodman, many other statisticians, the American Statistical Association, and I are concerned about the misuse of frequentist statistics and the misuse of the p value (*1*). Ronald Fischer would likely agree; he thought that a "significant" result required further experiments for confirmation. I think it time to go beyond frequentist statistics and integrate pretest likelihood, i.e. Bayesian thinking, into most statistical testing.

## INTRODUCTION

Imagine that you could calculate posttest probabilities, i.e. Bayes' theorem with simple addition. This paper shows how. To do so, we need to stop thinking of probabilities as ranging from 0 to 1.0 (i.e. 0 to 100%) and start thinking of probabilities as Weights (W) and Effect Size as Impact (I). W and I will be formally defined below, but it is best to first see how they work:

Suppose there exists a treatment that reduces heart attacks over a defined duration from 61% to 50%. The probability 61% corresponds to a $W_{prettreatment} = 0.2$ and a treatment effect from 61% to 50% corresponds to an effect size $I_{treatment} = –0.2$. Bayes' theorem is now calculated as $W_{posttreatment} = W_{prettreatment} + I_{treatment} = 0.2 + –0.2 = 0$ where W=0 corresponds to a 50% probability. This is Bayes' theorem with simple addition.

Now suppose there are six treatments that individually decrease heart attacks from 61% to 50% and suppose that these treatments are additive. If a person were to initiate all six treatments, some might think the benefit would be 6 times the 11% percentage point drop from 61% to 50% – this would be 61% minus 6 times 11% (66%) which would be negative 5% – this is clearly impossible because percentage change is not additive. With the linear form of Bayes' theorem, $W_{posttreatment} = W_{prettreatment}$ + 6 times $I_{treatment} = 0.2 + 6\ (–0.2) = –1.0$ which corresponds to a probability of 9%.

We can change $W_{pretreatment}$ and easily redo the calculations. If $W_{prettreatment} = –0.4$ (28%) and $I_{treatment} = –0.2$, for one treatment $W_{posttreatment} = W_{prettreatment} + I_{treatment} = –0.4 + –0.2 = –0.6$ (20%). For six treatments, $W_{posttreatment} = W_{prettreatment}$ + 6 times $I_{treatment} = –0.4 + 6\ (–0.2) = –1.6$ (2.5%). All that is required is that we think of probability as W values and effect size as I values.

In this example, W and I values are additive, i.e. they are part of a linear probability space. In contrast percentage change is not linearly additive. In a companion paper (Rembold, companion paper 2), I will discuss which types of data are already in a linear probability space and which types of data need transforms to be in a linear probability space. For the rest of this paper, I will only discuss probability data that exists between the endpoint of 0 and 1 (i.e. 0 to 100%). Importantly, this paper discusses only changes in probability because probabilities of two or more distinct things can be added – for example, you can die of heart disease or cancer or something else, the sum of heart disease, cancer, and something else deaths is 1 (100%).

I propose three statistics within linear probability space:

1) W (weight), a measure of the probability of something.

2) I (impact), something that changes a probability. When one object impacts another, it moves the second object. Therefore a test or treatment can "impact" the W (the probability) by the quantitative measure I. I values further from zero suggests a larger treatment effect and I values closer to zero suggest a smaller treatment effect. Positive I values increase W (probability) and negative I values reduce W (probability). I values are the same as what statisticians call Effect Size. In linear probability space Bayes' theorem reduces to $W_{posttest} = W_{pretest} + I$.

3) C (certainty) replaces "significance" to describe confidence about a result. C should be reported with the n value for the number of trials or tests done.

I recognize that many of these ideas are not new and were first discovered by the giants of statistical methods. They are well described (and buried) within textbooks such as Armitage & Berry (*5*). However, many scientists have ignored the idea of and the importance of analysis within a linear space. I apologize for not crediting the many statisticians and theoreticians on whose work this synthesis is built.

## INTRODUCTION TO CALCULATING IN LINEAR PROBABILTY SPACE: W VALUES

The naturally occurring linear probability space occurs when probability data are transformed to the common logarithm of the odds ratio ($\log_{10}$ odds) (*5*). Within this space, interpretation of data, statistical testing, and Bayes' theorem are linearly additive (Eqn. 1-3 (*6*)). The log of odds has been described as "weight of evidence (*7, 8*)," "Weight ("W") (*6*), and also as the logit transform (as ln odds, (*5*)-p. 394). Figure 1 shows the correspondence of W with traditional probabilities.

Some suggest using the natural log (ln) rather than the common log ($\log_{10}$) but as will be apparent later, the $\log_{10}$ is more intuitive, especially with C values. In my synthesis, W is a measure of probability.

For example, if probability of an event is 0.80 (80%), odds are 0.8/(1–0.8) = 0.8/0.2 = 4:1 = 4. So W = $\log_{10}$ 4 = 0.6. The conversions between probability, odds and W are:

$\text{Odds} = \frac{\text{Probability}}{1-\text{Probability}}$ In Excel: Odds = Probability / (1–Probability) (1)

$W = \text{Weight} = \log_{10} \text{Odds} = \log_{10} \frac{\text{Probability}}{1-\text{Probabilit}} = \log_{10} \frac{\text{Percentage}/100}{1-\text{Percentage}/100}$

In Excel: W = LOG10(Probability / (1–Probability)) (2)

$\text{Probability} = \frac{10^W}{1+10^W}$ In Excel: Probability=POWER(10,W)/(1+POWER(10,W)) (3)

| W (weight) | Probability | |
|---|---|---|
| 9 | 0.999 999 999 | |
| 6 | 0.999 999 | |
| 4 | 0.999 9 | |
| 3 | 0.999 | |
| 2 | 0.99 | |
| 1.3 | 0.952 | |
| 1 | 0.91 | |
| 0.6 | 0.80 | |
| 0.4 | 0.72 | |
| 0.2 | 0.61 | |
| 0.1 | 0.56 | if start as 0 (50%) and I is negative, |
| 0 | 0.50 | Low I: 0 to –0.09 (reduces from 50% to range 45% – 50%) |
| –0.1 | 0.44 | Intermediate I: –0.1 to –0.19 (reduces from 50% to 39% – 44%) |
| –0.2 | 0.39 | High I: –0.2 to –0.39 (reduces from 50% to range 29% – 38%) |
| –0.4 | 0.28 | Very high I: < –0.4 (reduces from 50% to 28% or lower) |
| –0.6 | 0.20 | |
| –1 | 0.09 | |
| –1.3 | 0.048 | C = –1.3 is a 51% chance a second identical study has a C<–1.3 |
| –2 | 0.01 | C = –2 is a 73% chance a second identical study has a C<–1.3 |
| –3 | 0.001 | |
| –4 | 0.000 1 | C = –4 is a 97% chance a second identical study has a C<–1.3 |
| –6 | 0.000 001 | |
| –9 | 0.000 000 001 | C = –9 is a 99.998% chance a second study has a C<–1.3 |

Figure 1. Correspondence of W (weight) and probability (multiply probability by 100 for percentage) for defined changes in W. Comments at right on I and C will be discussed below.

For optimal use of linear probability space, the user needs to familiarize her/himself with Fig. 1 more than the formulae above. W values replace probabilities. One advantage of W is that it has a more human scale: an average probability of 0.5 corresponds to a W = 0 and probabilities that range from 0.000 000 001 to 0.999 999 999 correspond to W values that range from –9 to 9.

In the midrange of W (–0.6 to +0.6), probability and W values are coarsely linear (i.e. a 0.2 change in W corresponding to an ~0.1 probability change). This can be remembered:

W of +0.6 is probability of 0.8 (80%),
W of +0.4 is probability of ~0.7 (~70%),
W of +0.2 is probability of ~0.6 (~60%),
W of 0 is probability of 0.5 (50%),
W of –0.2 is probability of ~0.4 (~40%),
W of –0.4 is probability of ~0.3 (~30%), and
W of –0.6 is probability of 0.2 (20%).

At high W values (>2), integer W values are the number of 9s in the probability, e.g. W=4 is probability 0.9999.

At low W values (<–2), integer W values are the number of 0s and the 1 in the probability, e.g. W=–3 is probability 0.001.

## EFFECT SIZE IS EASILY CALCULATED AND IS TERMED IMPACT

Effect size is a quantitative measure of how much means of two measures differ rather than how certain we are about the difference in means. A large advantage of linear probability space is that Effect Size is easily calculated as the difference of the means with a test or a treatment within linear probability space. I suggest naming effect size as Impact (I), where:

$$\text{Impact} = I = W_{mean2} - W_{mean1} \quad (4)$$

When one object impacts another, it moves the second object. Therefore a test or treatment can "impact" the W (weight, i.e. probability) by the quantitative measure I.

Since I moves a measurement from $mean_1$ to $mean_2$ in linear probability space, we need to consider the W of the mean data. For example, if $W_1 = 0$ (50%) and $W_2 = -0.2$ (39%), then $I = W_2 - W_1 = -0.2 - 0 = -0.2$. And if $W_1 = -0.95$ (10%) and $W_2 = -1.15$ (6.6%), then $I = -1.15 - -0.95 = -0.2$. Note that the difference is linear only in linear probability space but not as percentages.

Impact (I) here is identical to the Impact (I) described in Eqn. 11 and 18 below for the diagnostic benefit of a test with discrete data.

Ideally, just knowing I would be adequate as a measure of impact, i.e. Effect Size. For those that need words describing impact perhaps the best would be:

1) Low Impact with I values close to zero: between –0.1 & 0.1
   e.g. reduces from 50% to range 44% – 50% & reduces from 10% to range 8.2% – 10% (these are examples for a negative I which reduces W),
2) Intermediate Impact with I values between –0.2 & –0.1 or I between 0.1 & 0.2
   e.g. reduces from 50% to range 39% – 44% & reduces from 10% to range 6.6% – 8.1%,
3) High Impact with I values between –0.4 & –0.2 or I between 0.2 & 0.4
   e.g. reduces from 50% to range 28% – 39% & reduces from 10% to range 4.3% – 6.5%, &
4) Very High Impact with I values less than –0.4 or greater than 0.4
   e.g. reduces from 50% to 28% or lower & reduces from 10% to 4.2% or lower).

For the above examples of changes in percentage, a positive I would increase W. For data not easily transformed into linear probability space, Impact is harder to imagine because the mapping into linear probability space introduces degrees of freedom (DF) for the mapping function and any constants involved.

## CHANGING FROM BRIGHT LINE SIGNIFICANCE TO CERTAINTY

One of the earliest statistical comparison was Fishers t test. Since multiple other methods have been developed including comparison of multiple groups. In this paper, I will only consider comparison of two groups, but this method this can be trivially extended to multiple groups.

Since p values have been inappropriately used, are not linear, and low p values are sometimes inappropriately interpreted as higher importance, I suggest calculating a value termed Certainty or C from the p value in the same way W is calculated from probability:

$$C = \text{Certainty} = \log_{10}\frac{\text{p value}}{1-\text{p value}} \qquad \text{In Excel: C = LOG10(p value / (1–p value))} \qquad (5)$$

C replaces "significance." C is on the same scale as W (Fig. 1). C values are more linear and have a more human scale: we get impressed by p = 0.000 001 more than p = 0.001 more than p = 0.048, it is more appropriate to present these same levels of certainty as C values of –6, –3, and –1.3, respectively. An advantage of C is that all current methods that estimate p are easily converted to C values with Eqn. 5. The disadvantage of C is any complicated method to calculate p is retained in calculations of C. There are similarities between C and the Z score but C values go beyond the Z score.

Ideally, knowing just the C value would be adequate as a measure of certainty. For those that need words describing certainty, there should be a logical method. I do not favor choosing thresholds on whether the current study crosses a line of "significance." A more appropriate method should be based on whether the probability of a second identical study would be have a C < –1.28 corresponding to a p < 0.05. This concept has been termed p of p or p(theta) and is well described in excellent work of Goodman, Shao, and Chou (*2, 3*), but unfortunately is not as well-known as it should be. I propose the following words describing C (certainty):

1) Certainty Similar with C values > –0.4
   corresponds to p > 0.28 which is less than ~1 SE multiple (at high DF) and
   to a <19% chance that a second identical study would have a C<–1.3 (p<0.05)
   [this category is controversial and assumes that the assignment of groups are agnostic, i.e. there is no assigned control group. If this is not desired, then it would be included in group 2 – Certainty Indeterminate],
2) Certainty Indeterminate with C values –0.4 to –1.3
   corresponds to p between 0.28 and 0.05 which is ~1.1–2.0 SE multiples and
   to a 19% to 51% chance that a second identical study would have a C<–1.3,
3) Certainty Marginally Different with C values –1.3 to –2
   corresponds to p between 0.05 and 0.01 which is ~2.0–2.6 SE multiples and
   to a 51% to 73% chance that a second identical study would have a C<–1.3,
4) Certainty Different with C values –2 to –4,
   corresponds to p between 0.01 & 0.000 1 which is ~2.6–3.9 SE multiples and
   to a 73% to 97% chance that a second identical study would have a C<–1.3,
5) Certainty Very Different with C < –4,
   corresponds to p < 0.000 1 which is > ~4 SE multiples and
   a >97% chance that a second identical study would have a C<–1.3 (p<0.05), and
6) Certainty Very Different (Physics/Genetics level) with C < –9,
   corresponds to p < 0.000 000 001 which is > ~6 SE multiples (at high DF), and
   a >99.998% chance that a second identical study would have a C<–1.3 (p<0.05).

## REPEATABILITY IMPROVES CERTAINTY

There is a repeatability crisis in science. One of the major proponents and a tester of repeatability is Brian Nosek and his Reproducibility Project / Center for Open Science at the University of Virginia (*4*).

Some of the current repeatability crisis may be attributed to the traditional definition of significance as p<0.05 (C<–1.3) on the current dataset. As noted above, a p<0.05 (C<–1.3) on one dataset predicts only a 51% chance that a second identical study would have a p<0.05 (C<–1.3, see the excellent work of Goodman, Shao, and Chou (*2, 3*)). Interestingly, this change in statistical certainty was not responsible for much of the lack of repeatability in Nosek's analysis (*4*).

One possible threshold is to require a single study to have

1) a C < –2, i.e. a 73% chance that a second identical study would have a C<–1.3 or

2) a C < –4, i.e. a 97% chance that a second identical study would have a C<–1.3,

An alternative approach is to ask for two studies to both find C < –1.3.

In a companion paper (Rembold, companion paper 1), I show that C values are not strictly additive, there is an inherent variability with SD values ~1 when C values are added as compared to when C values are calculated from combined data. In general C values can be added but need a correction factor of adding +1 for each study added to the first and another +1 added if the data is not normally distributed. There is no need for correction if the data have different number of data points or different SD in the groups.

Multiple analyses of data, i.e. data mining or intermediate analysis, is a clear problem causing more issues with reproducibility. A solution is that for each additional test performed (either at the same time or sequentially ever), then the C value should be corrected by dividing it by the square root of the number of tests.

## CALCULATION OUTSIDE OF LINEAR PROBABLITY SPACE INDUCES ERROR

An important technical advantage of calculating in linear probability space is the elimination of calculation errors when mean and SD are calculated in any nonlinear space, which includes probabilities. To demonstrate this, I created a symmetric data set in percentage space with a mean of 50% and SD of 3.40037% (top left of Fig. 2). When the data are transformed into W and analyzed in linear probability space, the corresponding W is 0 and SD 0.059. When the mean and SD from W are transformed back into percentages, the mean is still 50% but the SD is slightly higher at 3.40870%, a small but potentially important difference.

When the data are divided by 2, the percentage space mean is 25% and SD 1.70018% (top right of Fig. 2). In linear probability space, the corresponding W is –0.4779 and SD 0.03952. When the mean and SD from W are transformed back into percentages, the mean is now 24.966% and SD 1.74317%, both different that calculations done in percentage space.

When the SD is higher (by doubling the data difference from 50%, bottom of Fig. 2), the differences are greater for both mean and SD.

These data show that when measurement variability is introduced, calculations of changes in probability are only accurate when done in a linear probability space, i.e. as weights.

| | Data with mean 0.5 | | | | Data divided by 2 - now mean ~0.25 | | | |
|---|---|---|---|---|---|---|---|---|
| | Percentage | Weight from Prob | Percentage from W | Difference | Percentage | Weight from Prob | Percentage from W | Difference |
| Data | 44.0 | -0.10 | | | 22.0 | -0.55 | | |
| | 47.0 | -0.05 | | | 23.5 | -0.51 | | |
| | 49.0 | -0.02 | | | 24.5 | -0.49 | | |
| | 49.5 | -0.01 | | | 24.8 | -0.48 | | |
| | 50.0 | 0.00 | | | 25.0 | -0.48 | | |
| | 50.5 | 0.01 | | | 25.3 | -0.47 | | |
| | 51.0 | 0.02 | | | 25.5 | -0.47 | | |
| | 53.0 | 0.05 | | | 26.5 | -0.44 | | |
| | 56.0 | 0.10 | | | 28.0 | -0.41 | | |
| Mean | 50.00000 | 0.00000 | 50.00000 | 0.00000 | 25.00000 | -0.47792 | 24.96553 | -0.03447 |
| SD | 3.40037 | 0.05931 | 3.40870 | 0.00833 | 1.70018 | 0.03952 | 1.74317 | 0.04298 |
| Data with double SD | 38.0 | -0.21 | | | 19.0 | -0.63 | | |
| | 44.0 | -0.10 | | | 22.0 | -0.55 | | |
| | 48.0 | -0.03 | | | 24.0 | -0.50 | | |
| | 49.0 | -0.02 | | | 24.5 | -0.49 | | |
| | 50.0 | 0.00 | | | 25.0 | -0.48 | | |
| | 51.0 | 0.02 | | | 25.5 | -0.47 | | |
| | 52.0 | 0.03 | | | 26.0 | -0.45 | | |
| | 56.0 | 0.10 | | | 28.0 | -0.41 | | |
| | 62.0 | 0.21 | | | 31.0 | -0.35 | | |
| Mean | 50.000 | 0.00 | 50.00000 | 0.00000 | 25.000 | -0.48 | 24.85946 | -0.14054 |
| SD | 6.80074 | 0.12009 | 6.86896 | 0.06822 | 3.40037 | 0.07991 | 3.59318 | 0.19281 |

Figure 2. Data showing the advantage of calculating means and standard deviations in a linear probability space (W) as compared to a percentage space (for probability divide by 100).

In a companion paper, I further evaluate the difference in the mean and SD calculations between percentage space and linear probability space (W) at various means. The error in SD is approximately one order of magnitude greater and in opposite direction as the error in the mean. The error in both is greater when the mean is further from W = 0, i.e. a probability of 50%. (Rembold, companion paper 2).

**AN EXAMPLE OF REAL CONTINUOUS DATA**

Figure 3 demonstrates statistical testing of a dataset easily converted to linear probability space. It shows measurements of the stoichiometric phosphorylation of a protein (myosin light chain) on a single residue (S19) in two datasets when carotid arteries were stimulated by high extracellular $K^+$.

These data are already probabilities (Fig. 4, 2nd and 3rd column) so are easily converted into in linear probability space by W = $\log_{10}$(probability/(1–probability)) (Fig. 3, 4th and 5th column). Then we calculate means ($x_1$ and $x_2$) and SD ($SD_1$ and $SD_2$) of W for each data set.

| | | | | | I | | | | | | | | C |
|---|---|---|---|---|---|---|---|---|---|---|---|---|---|
| | Raw Data | | Weight from Prob | | Impact | 95% CI of | | | | t | | | Certainty |
| | 0 min | 1 min | 0 min | 1 min | Dif means | Impact | | SD | SE | SE mult | DF | p (t distr) | W of p |
| Dataset 1 | 0.088 | 0.281 | -1.02 | -0.41 | | | | | | | | | |
| from Fig 7 | 0.079 | 0.467 | -1.07 | -0.06 | | | | | | | | | |
| | 0.186 | 0.472 | -0.64 | -0.05 | | | | | | | | | |
| | 0.091 | 0.471 | -1.00 | -0.05 | | | | | | | | | |
| | | | | | | | | | | | | | |
| Mean | 0.111 | 0.423 | -0.931 | -0.141 | 0.79 | 1.05 | 0.53 | 0.19 | 0.13 | 5.99 | 6 | 9.76E-04 | -3.01 |
| SD (stdev) | 0.050 | 0.095 | 0.195 | 0.178 | | | | | | | | | Different |
| n | 4 | 4 | 4 | 4 | | | | | | | | | |
| | | | | | | | | | | | | | |
| Dataset 2 | 0.240 | 0.480 | -0.50 | -0.03 | | | | | | | | | |
| from Fig 5 | 0.200 | 0.480 | -0.60 | -0.03 | | | | | | | | | |
| | 0.290 | 0.420 | -0.39 | -0.14 | | | | | | | | | |
| | 0.250 | 0.510 | -0.48 | 0.02 | | | | | | | | | |
| | | | | | | | | | | | | | |
| Mean | 0.245 | 0.473 | -0.492 | -0.048 | 0.44 | 0.55 | 0.34 | 0.08 | 0.05 | 8.09 | 6 | 1.91E-04 | -3.72 |
| SD (stdev) | 0.037 | 0.038 | 0.088 | 0.066 | | | | | | | | | Different |
| n | 4 | 4 | 4 | 4 | | | | | | | | | |
| | | | | | | | | | | | | | |
| Datasets 1&2 | | | | | | | | | | | | | |
| Mean | 0.178 | 0.448 | -0.712 | -0.095 | 0.62 | 0.83 | 0.41 | 0.22 | 0.11 | 5.74 | 12 | 9.37E-05 | -4.03 |
| SD (stdev) | 0.083 | 0.072 | 0.273 | 0.134 | | | | | | | | | Very Different |
| n | 8 | 8 | 8 | 8 | | | | | | | | | |
| | | | | | | | | | Sum C from Dataset 1 & 2 | | | | -6.73 |
| | | | | | | | | | | | | n | 2 |

Figure 3. Statistical testing in a dataset easily converted to linear probability space. The measurement is serine (S) 19 myosin light chain phosphorylation (MLCP) present in swine carotid arterial smooth muscle homogenates frozen prior to and 1 min after depolarization with 109 mM $K^+$ from two different data sets in the same paper (labeled dataset 1 and 2, (*9*)) Dilutions of samples containing myosin light chains were separated on isoelectric focusing gels and immunostained, so the stoichiometric ratio (phosphorylated as a percent of total myosin light chain) is reasonably accurate within the range 10–90% (*10*).

I (impact) = $x_2 - x_1$

Then we can calculate $SD_{x1\text{-}x2}$ for the W as $SD_{x1\text{-}x2} = \sqrt{\dfrac{(n_1-1)SD_1{}^2 + (n_2-1)SD_2{}^2}{(n_1-1)+(n_2-1)}}$ (6)

Then calculate $SE_{x1\text{-}x2} = \sqrt{SD_{x_1-x_2}{}^2\left\{\frac{1}{n_1}+\frac{1}{n_2}\right\}}$ (7)

The 95% Confidence intervals (95% CI) = I $\pm$ 1.96 · $SE_{x1\text{-}x2}$. 95% CI are for C<–1.3 (p<0.05).

t = SE multiples = $\dfrac{x_1-x_2}{SE_{x_1-x_2}} = \dfrac{I}{SE_{x_1-x_2}}$ (8)

From a t table or formula, p is calculated. In Excel p = tdist(t, DF, 2).

C = $\log_{10}\dfrac{\text{p value}}{1-\text{p value}}$

The datasets show with a C of –3.0 and –3.7 that the mean S19 myosin light chain phosphorylation increased with 1 min of $K^+$ depolarization in dataset$_1$ (top) and dataset$_2$ (bottom), respectively. In dataset$_1$, I = +0.8 corresponding to means increasing from 0.11 to 0.42 and in dataset$_2$ I = +0.44 corresponding to means increasing from 0.24 and 0.47. These both fit into the Very High impact (>+0.4) category.

Both datasets fit into the Different category which is where C is between –2 to –4 corresponding to a p value between 0.01 and 0.000 1 and for each the chance of second identical test having a C<–1.3 is 73-97% (in this case the second identical test had a C=–3.7). There were two independent experiments showing similar direction and C values, suggest higher confidence.

The sum of the Cs from datasets 1 and 2 individually was –6.7, a value higher than the C if is calculated when the datasets are analyzed together (C = –4.0, Fig. 3, bottom). This discrepancy partly is from chance and is partially real. When all the pairwise permutations (total of 8) of the above data were tested, the mean C1 was –2.1, the mean C2 was –2.9, the sum of the Cs from datasets 1 and 2 individually was –5.0, and the C when the datasets are analyzed together was –4.0). This suggests that when C values are added, they overestimate the C value by ~1 compared to when the data were analyzed together, this value of ~1 agrees with a more complete evaluation of the additivity of C (Rembold, companion paper 1).

A spreadsheet for these calculations is included in the appendix.

In another publication, I will present how to do calculations on data that are not easily converted into linear probability space (Rembold, companion paper 2).

## ANALYSIS OF DISCRETE DATA

Some data presents as counts, i.e. as integers. When converted to rates, these data are in regular probability space. An example of these data would be recurrence of cancer or cardiac ischemia after an intervention. Statistics can be generated once a large number are measured. These data are best presented as a 2x2 table:

| | | Outcome/Disease | | | | |
|---|---|---|---|---|---|---|
| | | Yes | No | Total | Rate (probability) | |
| Factor/Test | Yes | a | c | a+c | $\frac{a}{a+c}$ | With Factor |
| | No | b | d | b+d | $\frac{b}{b+d}$ | Without Factor |
| | Total | a+b | c+d | a+b+c+d = n | | |
| | | With Outcome | Without Outcome | | | |

True positive rate $\frac{a}{a+b}$ = sensitivity

False negative rate $\frac{b}{a+b}$ = 1 – sensitivity

False positive rate $\frac{c}{c+d}$ = 1 – specificity

True negative rate $\frac{d}{c+d}$ = specificity (9)

From the can be calculated several statistics with varying advantages and disadvantages:

| | Additive | Time Dependent | Baseline Risk Included | |
|---|---|---|---|---|
| Odds ratio (OR) = $\frac{ratio\ with\ factor}{ratio\ without\ factor} = \frac{a/c}{b/d} = \frac{ad}{bc}$ | No | No | No | (10) |
| I (impact) = $\log_{10}$ OR = $\log_{10} \frac{ad}{bc}$ | **Yes** | No | No | (11) |
| Relative risk (RR) = $\frac{rate\ with\ factor}{rate\ witho\ \ \ factor} = \frac{a/(a+c)}{b/(b+d)} = \frac{a(b+d)}{b(a+c)}$ | No | No | No | (12) |
| Absolute Risk Reduction (ARR) = $rate_+ - rate_- = \frac{b}{b+d} - \frac{a}{a+c}$ | No | Yes | Yes | (13) |
| Number needed to treat (NNT) = 1 / ARR | No | Yes | Yes | (14) |

The odds ratio (OR, Eqn. 10) is not time dependent (assuming linear risk), does not include baseline risk, and is not additive. It has been called approximate relative risk (this is only valid if a << c and b << d, i.e. a low rate).

The impact (I, Eqn. 11) of a treatment is also not time dependent (assuming linear risk), does not include baseline risk, but importantly, it is linearly additive. Note that this I is the same as I described above in Eqn. 4 for continuous data.

Relative risk (RR, Eqn. 12) is not time dependent (assuming linear risk), does not include baseline risk, and is not additive.

Relative risk reduction (RRR) = 1 – RR is not time dependent (assuming linear risk), does not include baseline risk, and is also not additive. The RRR is the most common statistic quoted by clinicians to patients to describe a treatment benefit.

Absolute Risk Reduction (ARR, eqn. 13) includes baseline risk, is time dependent and is not additive.

Number needed to treat (NNT, eqn. 14) and number needed to screen (NNS, (*11*)) include baseline risk, are time dependent and are not additive

In summary, the only additive statistic is I. ARR, NNT, NNS include baseline risk and are time dependent. RR, RRR, and I do not include baseline risk and are not time dependent (assuming linear benefit of treatment).

## INTERPRETATION OF TEST RESULTS

A clear advantage of calculating in linear probability space is the evaluation of test results. The $W_{pretest}$ (pretest probability) can be added to the I of a test result ($I_+$ and $I_-$ for positive and negative tests, respectively, Eqn 15 & 16) to create a $W_{posttest}$ (posttest probability, Eqn. 17).

Calculation of I values for test results (*6*) is done as follows:

I of a Positive test: $I_{+test} = \log\frac{Sensitivity}{1-Specificit} = \log\frac{True\ Positive\ Rate}{False\ Positive\ Rate} = \log\frac{a/(a+b)}{c/(c+d)} = \log\frac{a(c+d)}{c(a+b)}$ (15)

I of a Negative test: $I_{-test} = \log\frac{1-Sensitivity}{Specificity} = \log\frac{False\ Negative\ Rate}{True\ Negative\ Rate} = \log\frac{b/(a+b)}{d/(c+d)} = \log\frac{b(c+d)}{d(a+b)}$ (16)

$W_{posttest} = W_{pretest} + I_{test\ 1} + I_{test\ 2} + \ldots$ (17)

For example, the prevalence of coronary artery disease (CAD) is 4% in the general population (*6*) so, $W_{pretest}$ is –1.38.

To this can be added I = –0.09 for a male aged 40-49, so $W_{posttest}$ = –1.38 – 0.09 = –1.47 (a 3% chance of CAD, see Fig. 1 to go back and forth from W to Probability).

Then add I = +1.39 for atypical chest pain (i.e., chest pain with some but not all features of CAD), so $W_{posttest}$ equals –1.47 + 1.39 = –0.08 (a 45% chance of CAD). Since $W_{posttest}$ = –0.08, a test to decide if the chest pain were CAD would be helpful.

Adding +1.39 $I_{+test}$ for a concordant positive stress ECG and positive nuclear scan increases $W_{posttest}$ to –0.08 + 1.39 = +1.31 (a 95% chance of CAD), suggesting the presence of CAD.

In contrast, adding –1.13 $I_{-test}$ for a concordant negative stress ECG and negative nuclear scan decreases $W_{posttest}$ to –0.08 – 1.13 = –1.21 (a 6% chance of CAD), suggesting the absence of CAD.

This example shows how stress nuclear testing can be very helpful for a clinician to diagnose the presence or absence of CAD in a 40-49 year old male with atypical chest pain.

Since W and I values are additive, the difference of I values of a positive ($I_{+test}$) and negative ($I_{-test}$) tests reveals the relative impact of the overall test ($I_{test}$).

$$I_{test} = I_{+test} - I_{-test} = \log_{10} \frac{Sensitivity \cdot Specificity}{(1-Specificity)(1-Sensitivit\ \ )} = \log_{10} \frac{\frac{a}{a+b}\,\frac{d}{c+d}}{\frac{c}{c+d}\,\frac{b}{a+b}} = \log_{10} \frac{ad}{bc} = \log_{10} \mathrm{OR} \quad (18)$$

For example, a stress ECG alone has an $I_{test}$ = +0.65 – –0.39 = +1.04, a stress nuclear scan an $I_{test}$ = +0.74 – –0.74 = +1.48, and a combined concordant stress ECG and nuclear has an $I_{test}$ = +1.39 – –1.13 = +2.52. These suggest that a concordant stress ECG and nuclear is better ($I_{test}$=+2.5) than a nuclear alone ($I_{test}$=+1.5) which is better than a stress ECG alone ($I_{test}$=+1.0). Overall, I propose that a test with a higher $I_{test}$ indicate that the test is better than a lower $I_{test}$.

Note that these $I_{test}$ values are independent of pretest probability. Since I values are linear, it can provide information along a whole receiver operated curve (ROC) if the test accuracy does not depend on pretest probability. Interestingly, $I_{test}$ is the log OR for the test.

## AN EXAMPLE OF DISCRETE DATA

Figure 4 demonstrates testing of discrete data with integer numbers of events in different groups. It shows the number of people with known atherosclerotic vascular disease who had a nonfatal myocardial infarction (MI) or cardiovascular death (CVD), i.e. events (a or b) or did not not have events (c or d) based on assignment to placebo (b or d) or treatment (a or c) in three different trials of statins (LDL cholesterol lowering agents). Specifically the trials were:

1) Placebo (control) or simvastatin 20-40 mg (treated) in the **4S** (Scandinavian Simvastatin Survival Study (*12*),

2) Placebo (control) or simvastatin 40 mg (treated) in the **HPS** (Heart Protection Study(*13*),

& 3) Simvastatin 20 (control, low intensity) or atorvastatin 80 mg (treated, high intensity) in the **IDEAL** trial (Incremental decrease in endpoints through aggressive lipid lowering (*14*))

| | | | | 4S | | HPS | | IDEAL |
|---|---|---|---|---|---|---|---|---|
| | Control | Treated | Control | Treated | Control | Treated | Control | Treated |
| Events | b | a | 691 | 464 | 1212 | 898 | 456 | 401 |
| No Events | d | c | 1532 | 1757 | 9055 | 9371 | 3993 | 4038 |
| Sum | b+d | a+c | 2223 | 2221 | 10267 | 10269 | 4449 | 4439 |
| Odds Ratio (OR) | | ad/bc | | 0.5855 | | 0.71594 | | 0.86959 |
| I = Impact | | log10 ad/bc = log10 OR | | -0.2325 | | -0.1451 | | -0.0607 |
| SE[ln OR] | | sqrt(1/a+1/b+1/c+1/d) | | 0.06946 | | 0.04643 | | 0.07201 |
| SE[log10 OR] | | SE[ln OR]/ln(10) | | 0.03016 | | 0.02016 | | 0.03127 |
| Impact 95%+ CI | | Impact - 1.96 SE(log10 OR) | | -0.2916 | | -0.1846 | | -0.122 |
| Impact 95%- CI | | Impact + 1.96 SE(log10 OR) | | -0.1734 | | -0.1056 | | 0.00061 |
| SE multiples = t | | \| Impact \| / SE(log10 OR) | | 7.7 | | 7.2 | | 1.9 |
| p (t distribution) | | tdist | | 1.58E-14 | | 6.37E-13 | | 5.23E-02 |
| C = Certainty (t) | | W of p = log10(p/(1-p)) | | -13.801 | | -12.196 | | -1.2579 |
| Chi square | n*(ad-bc)^2/((a+b)(c+d)(a+c)(b+d) | | | 60.0 | | 52.1 | | 3.8 |
| p (chi square) | | tdist | | 9.46E-15 | | 5.15E-13 | | 5.22E-02 |
| C = Certainty (Chi sq) | | W of p = log10(p/(1-p)) | | -14.024 | | -12.288 | | -1.2594 |
| Control rate | | b/(b+d) | | 0.311 | | 0.118 | | 0.102 |
| Treated rate | | a/(a+c) | | 0.209 | | 0.087 | | 0.090 |
| ARR | | b/(b+d)-a/(a+c)= cont-treated | | 0.102 | | 0.031 | | 0.012 |
| RR | | a(b+d) / b(a+c) | | 0.672 | | 0.741 | | 0.881 |
| RRR | | 1 - RR | | 0.328 | | 0.259 | | 0.119 |
| Weight control | | log(control rate/(1-control rate) | | -0.3458 | | -0.8734 | | -0.9423 |
| Weight treated | | log(treated rate/(1-treated rate) | | -0.5783 | | -1.0185 | | -1.003 |
| I = Impact (from ARR) | | Wtreated - Wcontrol | | -0.2325 | | -0.1451 | | -0.0607 |
| LDL at start | | | | 190 | | 131 | | 104 |
| Duration (years) | | | | 5.4 | | 5 | | 6.7 |
| Tau (exponetial fit in y) | | tau=-t/ln(1-control rate) | | 14.51 | | 39.80 | | 61.96 |
| Control rate 10 years | | 1-exp(-t/tau) | | 0.498 | | 0.222 | | 0.149 |
| Weight cont 10 years | | log10(cont rate/(1-cont rate)) | | -0.003 | | -0.544 | | -0.757 |
| Weight treat | | Weight cont + Weight ARR | | -0.236 | | -0.689 | | -0.817 |
| Treated rate | | 10^W/(1+10^W) | | 0.368 | | 0.170 | | 0.132 |
| ARR 10 years | | Rate control - Rate treated | | 0.131 | | 0.052 | | 0.017 |
| Control rate 20 years | | 1-exp(-t/tau) | | 0.748 | | 0.395 | | 0.276 |
| Weight cont 20 years | | log10(cont rate/(1-cont rate)) | | 0.473 | | -0.185 | | -0.419 |
| Weight treat | | Weight cont + Weight ARR | | 0.240 | | -0.330 | | -0.480 |
| Treated rate | | 10^W/(1+10^W) | | 0.635 | | 0.319 | | 0.249 |
| ARR 20 years | | Rate control - Rate treated | | 0.113 | | 0.076 | | 0.027 |
| Maximal ARR | | | | 0.133 | | 0.083 | | 0.035 |
| occurs at a control rate of | | | | 0.57 | | 0.54 | | 0.52 |
| Maximal NNT | | NNT = 1/ARR | | 8 | | 12 | | 29 |
| Rate -I/2 (control) | | 10^(-I/2) / ( 1 + 10^(-I/2) ) | | 0.567 | | 0.542 | | 0.517 |
| Rate I/2 (treated) | | 10^(I/2) / ( 1 + 10^(I/2) ) | | 0.433 | | 0.458 | | 0.483 |
| ARR maximal | | Rate control - Rate treated | | 0.133 | | 0.083 | | 0.035 |
| NNT maximal | | 1/ARR | | 7.517 | | 11.998 | | 28.637 |

Figure 4. Statistical testing in three datasets with numbers of events in different groups – specifically treating people with atherosclerotic coronary artery disease with statins. Trial names are described in the text.

OR equal ad/bc.

I = Impact = $\log_{10}$ ad/bc = $\log_{10}$ OR.

$SE_{\ln OR} = \sqrt{1/a + 1/b + 1/c + 1/d}$ from (*5*). (19)

$SE_I = SE_{\log 10\ OR} = SE_{\ln OR} / \ln(10)$ (20)

The 95% Confidence intervals (95% CI) = I ± 1.96 · $SE_I$. This is for C<–1.3 (p<0.05).

t = SE multiples = |I| / $SE_I$.

p value is from t table or in Excel p value = tdist(t, DF, 2)

$C = \log_{10} \frac{\text{p value}}{1 - \text{p value}}$

ARR was then calculated as (b/(b+d)–a/(a+c)).

In this example, I values were –0.23 (high impact), –0.15 (intermediate impact), and –0.06 (low impact), for the 4S, HPS, and IDEAL respectively. These I values marginally correlated with the LDL cholesterol concentrations prior to treatment of 190, 131, and 104 mg $dl^{-1}$ ($r^2$ = 0.96, $C_{regression}$ = –1.39, marginally different), respectively suggesting a greater benefit of statin treatment at a higher LDL cholesterol concentrations (*15, 16*). I values were the same when calculated as $\log_{10}$ ad/bc or as the ARR based $W_{treated} - W_{control}$.

The C values were –14, –12 and –1.26, respectively suggesting very high certainty for the 4S and HPS results (very different at physics/genetics level) and very low certainty in the IDEAL trial (indeterminate). Similar p values and C were found with chi square calculations and with t testing.

The ARR were 0.10, 0.030, and 0.012 for the 4S, HPS, and IDEAL, respectively. These ARR correlated well with the LDL cholesterol concentrations prior to treatment ($r^2$ = 0.99, $C_{regression}$ = –1.87, marginally different).

To calculate C from r values and n, use the formula from Armitage and Berry (*1*):

$$t_{regression} = \sqrt{\frac{(n-2)\, r^2}{1 - r^2}} \quad (21)$$

Then calculate $C_{regression}$ to be calculated from p which is calculate from $t_{regression}$ and DF.

I suggest that the I and C values presented here are an alternate answer to Prof. Goodman's question about how to present clinical trial results. The I, i.e. the clinical impact, of the 4S trial (statins vs. placebo with high LDL cholesterol, –0.23) is 50% higher than the I of the HPS trial (statins vs placebo with intermediate LDL cholesterol, –0.15) and nearly fourfold greater than the I of the IDEAL trial (low vs high intensity statin treatment, –0.06) suggesting impact for 4S > HSP > IDEAL.

In the case of statins it is possible and reasonable to propose that the treatment benefit persists with longer treatment. If we assume that the benefits of statins are linear over time and we account for the declining numbers as people have myocardial infarctions or cardiovascular death (by calculating the exponential curve fit constant tau = –duration/ln(1-control rate)), then we can calculated ARR for specific durations of therapy. In the 4S trial, ARR would be 0.13 for 10, years and 0.11 for 20 years – the decline at 20 years is caused by the high rate of events in the control and treated groups that deplete the number of patients. By calculating ARR at all control rates and finding the $ARR_{maximal}$, the $NNT_{maximal}$ for each trial was calculated to be 8, 12, and 29 for the 4S, HPS, and IDEAL, respectively.

The $NNT_{maximal}$ can also be directly calculated from only I values. The largest effect of a negative I values (for a reduction) occurs when W is reduced from –I/2 to I/2, i.e. values surrounding W=0 (a 50% probability).

Therefore $NNT_{maximal} = \frac{1}{\text{ARRmaximal}} = 1\Big/\left( \frac{10^{-I/2}}{1+ 10^{-I/2}} - \frac{10^{I/2}}{1+ 10^{I/2}} \right)$.

So for the 4S trial where I = —0.2325,

$NNT_{maximal} = 1\Big/\left( \frac{10^{0.2325/2}}{1+ 10^{0.2325/2}} - \frac{10^{-0.2325/2}}{1+ 10^{-0.2325/2}} \right)$ = 1/(0.567 – 0.433) = 1/0.133 = 7.5.

A spreadsheet for these calculations is included in the appendix.

Figure 5 analyzes the datasets from Goodman's 2016 Science paper (*1*): all three datasets have the relatively macabre control event rate of 80% which is reduced to 74% with n=900 in trial 1, to 60% with n=100 in trial 2, and to 74% with n=500 in trial 3. The I values were –0.15, –0.43, and –0.15, suggesting a threefold greater impact in trial #2 (very high impact) than trial #1 and #3 (intermediate impact). The C values were –1.5, –1.5 and –0.9, suggesting a marginally different certainty for trial #1 and #2 and indeterminate certainty for trial #3, however, the C values were similar for the three trials.

| | | | | Goodman 1 | | Goodman 2 | | Goodman 3 | |
|---|---|---|---|---|---|---|---|---|---|
| | Control | Treated | | Control | Treated | Control | Treated | Control | Treated |
| Events | b | a | | **360** | **333** | **40** | **30** | **200** | **185** |
| No Events | d | c | | **90** | **117** | **10** | **20** | **50** | **65** |
| Sum | b+d | a+c | | **450** | **450** | **50** | **50** | **250** | **250** |
| **Odds Ratio (OR)** | | **ad/bc** | | | **0.71154** | | **0.375** | | **0.71154** |
| **I = Impact** | | **log10 ad/bc = log10 OR** | | | **-0.1478** | | **-0.4260** | | **-0.1478** |
| **SE[ln OR]** | | **sqrt(1/a+1/b+1/c+1/d)** | | | **0.1595** | | **0.45644** | | **0.21399** |
| **SE[log10 OR]** | | **SE[ln OR]/ln(10)** | | | **0.06927** | | **0.19823** | | **0.09293** |
| **Impact 95%+ CI** | | **Impact - 1.96 SE(log10 OR)** | | | **-0.2836** | | **-0.8145** | | **-0.33** |
| **Impact 95%- CI** | | **Impact + 1.96 SE(log10 OR)** | | | **-0.012** | | **-0.0374** | | **0.03435** |
| **SE multiples = t** | | **\| Impact \| / SE(log10 OR)** | | | **2.1** | | **2.1** | | **1.6** |
| p (t distribution) | | tdist | | | 0.0331 | | 0.0341 | | 0.1124 |
| **C = Certainty (t)** | | **W of p = log10(p/(1-p))** | | | **-1.4651** | | **-1.4524** | | **-0.8976** |
| Chi square | n*(ad-bc)^2/((a+b)(c+d)(a+c)(b+d) | | | | 4.6 | | 4.8 | | 2.5 |
| p (chi square) | | tdist | | | 0.0325 | | 0.0291 | | 0.1109 |
| **C = Certainty (Chi sq)** | | **W of p = log10(p/(1-p))** | | | **-1.4742** | | **-1.5233** | | **-0.9039** |
| | | | | | | | | | |
| Control rate | | b/(b+d) | | | 0.800 | | 0.800 | | 0.800 |
| Treated rate | | a/(a+c) | | | 0.740 | | 0.600 | | 0.740 |
| **ARR** | | **b/(b+d)-a/(a+c) =cont-treated** | | | **0.060** | | **0.200** | | **0.060** |
| RR | | a(b+d) / b(a+c) | | | 0.925 | | 0.750 | | 0.925 |
| RRR | | 1 - RR | | | 0.075 | | 0.250 | | 0.075 |
| | | | | | | | | | |
| Weight control | | log(control rate/(1-control rate) | | | 0.60206 | | 0.60206 | | 0.60206 |
| Weight treated | | log(treated rate/(1-treated rate) | | | 0.45426 | | 0.17609 | | 0.45426 |
| **I = Impact (from ARR)** | | **Wtreated - Wcontrol** | | | **-0.1478** | | **-0.426** | | **-0.1478** |

Figure 5. Statistical testing in a three datasets from Goodman's 2016 Science paper (*4*).

Finally, I analyzed a number of clinical trials for prevention of cardiovascular (CV) events (defined as myocardial infarction (MI) and CV death for multiple treatments including lipid lowering, diet, anti-inflammatory, and chelation that were done in people 1) without known atherosclerosis (primary prevention), 2) with inflammation but without known atherosclerosis (primary+hsCRP), 3) with known atherosclerosis (secondary prevention), 4) with an acute coronary syndrome (ACS), and 5) with end stage renal disease (ESRD). These results were sorted by impact (I) from very high to low. Also presented are number of trials and C values with classifications of certainty. Number of subjects and events (outcomes) were combined for multiple trials and then I and C values were calculated. I did not include trials in which allowed physicians to add or increase the dose of other lipid lowing agents to people in the control and/or the treatment groups.

There were different to very different C values (<–2) seen in treatments with various I values (Fig. 6):

A very high I was seen in ACS and secondary prevention with the Mediterranean diet.

High I was seen in inflammation without known atherosclerosis with statins, in secondary prevention with partial ileal bypass, and in primary prevention with the Mediterranean diet (CVA was included in the outcome for this trial).

Intermediate I was seen in primary prevention with statins and in secondary prevention with statins, omega 3 fatty acids (EPA) when triglycerides were 150-500 mg $dl^{-1}$, the fibrate gemfibrozil, & niacin.

Low I was seen in ACS with a PCSK9 inhibitor (alirocumab) and in secondary prevention with a PCSK9 inhibitor (evolocumab), a high vs. low intensity statin, ezetimibe added to a statin, & the CETP blocker anacetrapib added to a statin. The most negative C values were seen in primary and secondary prevention with statins given the larger number of trials performed.

There were marginally different C values (–1.3 to –2) in treatments with lower I values:

Intermediate I was seen in primary prevention with the fibrate gemfibrozil.

Low I was seen with secondary prevention with an anti-inflammatory agent (canakinumab) and in ESRD with statins.

There were indeterminate C values (–0.4 to –1.3) in treatments with lower I values: primary prevention with cholestyramine, secondary prevention with chelation and in ACS with statins.

These data should allow the clinician to prioritize anti-atherogenic therapy. Clearly the Mediterranean diet reduced CV events. Lowering LDL cholesterol with statins reduced CV events in primary and secondary prevention, but other LDL lowering agents such as PCSK9 blockers, ezetimibe, and partial ileal bypass also reduced CV events in secondary prevention. Statins had lower impact in ACS and ESRD than in primary or secondary prevention. The fibrate gemfibrozil and if triglycerides were elevated, high dose omega 3 fatty acids reduced CV events in secondary prevention.

| | Treatment | Population | LDL | I = Impact | 95% CI | n | C =Certainty | Call | Trials |
|---|---|---|---|---|---|---|---|---|---|
| **Very high Impact (moves 50% to <29%)** | | | | | | | | | |
| | Mediterranean Diet | ACS | 176 | -0.55 | -0.91 to -0.17 | 1 | -2.3 | Different | Lyon |
| | Mediterranean Diet | Secondary | 176 | -0.50 | -0.87 to -0.12 | 1 | -2.0 | Different | Lyon |
| **High Impact (moves 50% to 29-38%)** | | | | | | | | | |
| | Statin | Primary+hsCRP | 108 | -0.27 | -0.42 to -0.12 | 1 | -3.4 | Different | JUPITER |
| | Partial Ileal Bypass | Secondary | 166 | -0.25 | -0.38 to -0.11 | 1 | -3.3 | Different | POSCH |
| | Mediterranean Diet | Primary | 132 | -0.20 | -0.29 to -0.10 | 2 | -4.3 | Very Different | EVOO and Nut arm of PREDIMED (with CVA) |
| **Intermediate Impact (moves 50% to 39-44%)** | | | | | | | | | |
| | Gemfibrozil | Primary | 188 | -0.19 | -0.34 to -0.04 | 1 | -1.8 | Marginally Different | HHS |
| | **Statin** | **Primary** | **152** | **-0.15** | **-0.19 to -0.10** | **5** | **-9.9** | **Very Different** | WOS, AFCAPS, PROSPER, ASCOT, MEGA |
| | **Statin** | **Secondary** | **143** | **-0.15** | **-0.18 to -0.13** | **4** | **-29.9** | **Very Different** | 4S, LIPID, CARE, HPS |
| | EPA (omega 3 fish oil) | Secondary | 75 | -0.13 | -0.19 to -0.06 | 1 | -4.3 | Very Different | Reduce-It |
| | Gemfibrozil | Secondary | 112 | -0.12 | -0.21 to -0.04 | 1 | -2.3 | Different | HIT |
| | Niacin | Secondary | 170 | -0.10 | -0.16 to -0.03 | 1 | -2.2 | Different | CDP-niacin arm |
| **Low Impact (moves 50% to 45-49%)** | | | | | | | | | |
| | *Cholestryamine* | *Primary* | *200* | *-0.09* | *-0.19 to +0.005* | *1* | *-1.2* | *Indeterminant* | *LRC* |
| | Evolocumab-statin vs statin | Secondary | 92 | -0.09 | -0.14 to -0.05 | 1 | -4.3 | Very Different | FOURIER (PSCK9) |
| | *Chelation* | *Secondary* | *90* | *-0.09* | *-0.20 to +0.02* | *1* | *-1.0* | *Indeterminant* | TACT |
| | Canakinumab | Secondary | 83 | -0.07 | -0.12 to -0.007 | 1 | -1.5 | Marginally Different | CANTOS |
| | **Statin high vs low intensity** | **Secondary** | **102** | **-0.06** | **-0.09 to -0.03** | **3** | **-3.6** | **Different** | IDEAL, TNT, SEARCH |
| | Alirocumab-statin vs statin | ACS | 92 | -0.06 | -0.10 to -0.02 | 1 | -2.2 | Different | Odyssey (PSCK9) |
| | *Statin* | *ACS* | *86* | *-0.06* | *-0.12 to +0.003* | *2* | *-1.2* | *Indeterminant* | PROVE-IT, AZ |
| | Ezetimibe-statin vs statin | Secondary | 70 | -0.06 | -0.09 to -0.02 | 1 | -2.9 | Different | IMPROVE-IT |
| | **Statin** | **ESRD** | **107** | **-0.05** | **-0.09 to -0.002** | **3** | **-1.4** | **Marginally Different** | 4D, AURORA, SHARP |
| | Anacetrapib-statin vs statin | Secondary | 61 | -0.05 | -0.09 to -0.01 | 1 | -2.1 | Different | REVEAL |

Figure 6. Clinical trial for prevention of myocardial infarction and cardiovascular death with varying treatments including lipid lowering, diet, anti-inflammatory, and chelation in various populations including primary prevention (without known CAD), secondary prevention (with known CAD, but no recent ACS), those with a recent acute coronary syndrome (ACS), those with end stage renal disease (ESRD) and primary prevention with inflammation as measured by high high-sensitivity C reactive protein (hsCRP). These results were sorted by I and also presented is C (*italics if marginal or indeterminate certainty*) and **bolded if more than 2 trials**).

Trial references are Lyon (*17, 18*), Jupiter (*19*), POSCH (*20*), PREDIMED (*21*), HHS (*22*), WOS (*23*), AFCAPS (*24*), PROSPER (*25*), ASCOT (*26*), MEGA (*27*), 4S (*12*), LIPID (*28*), CARE (*29*), HPS (*13*), Reduce-It (*30*), HIT (*31*), CDP (*32, 33*), LRC (*34*), FOURIER (*35*), TACT (*36*), CANTOS (*37*), IDEAL (*14*), TNT (*38*), SEARCH (*39*), Odyssey (*40*) PROVE-IT (*41*), AZ (*42*), IMPROVE-IT (*43*), 4D (*44*), AURORA (*45*), SHARP (*46*), and REVEAL (*47*). The outcome for PREDIMED are the 2018 data, includes stroke, and are corrected for person years. The CDP (*32, 33*) trial compared niacin alone vs. placebo in the pre statin era.